\documentclass[preprint]{aastex}

\newcommand{\HST}{{\sl HST}}

\newcommand{\Msun}{\mbox{$M_{\sun}$}}

\newcommand{\Lsun}{\mbox{$L_{\sun}$}}

\newcommand{\Mjup}{\mbox{$M_{Jup}$}}

\newcommand{\etal}{et al.}
\newcommand{\eg}{e.g.}
\newcommand{\ie}{i.e.}

\newcommand{\kms}{\hbox{km~s$^{-1}$}}

\newcommand{\Kp}{\mbox{$K^{\prime}$}}
\newcommand{\Ks}{\mbox{$K_S$}}
\newcommand{\degs}{\mbox{$^{\circ}$}}
\newcommand{\Lbol}{\mbox{$L_{bol}$}}
\newcommand{\Teff}{\mbox{$T_{\rm eff}$}}

\slugcomment{Submitted to ApJ June 2, 2005; accepted August 1, 2005}

\shorttitle{Kelu-1 is a Binary L Dwarf}
\shortauthors{Liu \& Leggett}

\begin{document}

\title{Kelu-1 is a Binary L Dwarf:\\ 
First Brown Dwarf Science from Laser Guide Star Adaptive Optics}


\author{Michael C. Liu\altaffilmark{1}}
\affil{Institute for Astronomy, University of Hawai`i, 2680 Woodlawn
Drive, Honolulu, HI 96822} 
\email{mliu@ifa.hawaii.edu}

\and

\author{Sandy K. Leggett}
\affil{United Kingdom Infrared Telescope, Joint Astronomy Centre, 660
  North A'ohoku Place, Hilo, HI 96720}
\email{skl@jach.hawaii.edu}

\altaffiltext{1}{Alfred P. Sloan Research Fellow} 


\begin{abstract}
We present near-infrared (1--2.4~\micron) imaging of the L~dwarf Kelu-1
obtained with the Keck sodium laser guide star adaptive optics (LGS AO)
system as part of a high angular resolution survey for substellar
binaries.  Kelu-1 was one of the first free-floating L~dwarfs
identified, and the origin of its overluminosity compared to objects of
similar spectral type has been a long-standing question.  Our images
clearly resolve Kelu-1 into a 0.29\arcsec\ (5.4~AU) binary with
near-infrared flux ratios of $\approx$0.5~mags.  A previous
non-detection of binarity by {\sl Hubble Space Telescope} demonstrates
that the system is a true physical pair and that its projected orbital
motion has been significant over the last 7~years.  Binarity explains
the properties of Kelu-1 that were previously noted to be anomalous
compared to other early-L~dwarfs. We estimate spectral types of L1.5--L3
and L3--L4.5 for the two components, giving model-derived masses of
0.05--0.07~\Msun\ and 0.045--0.065~\Msun\ for an estimated age of
0.3--0.8~Gyr.  More distant companions are not detected to a limit of
$\approx$5--9~\Mjup.
The presence of Li~6708~\AA\ absorption indicates that both components
are substellar, but the weakness of this feature relative to other
L~dwarfs can be explained if only Kelu-1B is Li-bearing.  Determining
whether both or only one of the components possesses lithium could
constrain the age of Kelu-1 (and other Li-bearing L~binaries) with
higher precision than is possible for most ultracool field objects.
These results are the first LGS AO observations of brown
dwarfs and demonstrate the potential of this new instrumental capability
for substellar astronomy.
\end{abstract}



\section{Introduction}

Over a decade since they were first discovered, brown dwarfs are now
being found in abundance from optical and infrared (IR) imaging surveys
\citep[e.g.][]{1999ApJ...519..802K, 1999A&AS..135...41D,
  2002astro.ph..4065H}.  While many hundreds of substellar objects have
been identified, their physical properties and their origin(s) continue
to be active areas of inquiry.  Multiplicity is a key pathway towards
understanding these issues.  The substellar binary frequency, separation
distribution, and mass ratio distribution can constrain the formation
mechanisms for brown dwarfs \citep[e.g.][]{2002MNRAS.332L..65B}.
Substellar binaries also provide systems with a common age and
metallicity, which can aid interpretation of physical properties such as
colors and spectra.  Finally, dynamical mass determinations for
substellar binaries are sorely needed to test the theoretical models
over a wide range of parameter space.  To date, only two ultracool
binaries have such measurements (GL~569Bab:
\citealp{2001ApJ...560..390L, 2004astro.ph..7334O}; 2MASSW
J0746425+2000321AB: \citealp{2004A&A...423..341B}).

Previous imaging surveys of nearby ultracool objects, primarily by {\sl
  Hubble Space Telescope} (\HST) at optical wavelengths
\citep{1999ApJ...526L..25K, 2001AJ....121..489R, 2003ApJ...587..407C,
  2003AJ....126.1526B, 2003AJ....125.3302G}, find that substellar
binaries typically have $\lesssim$0.5\arcsec\ separations and that
smaller separations are more common than larger ones --- therefore, high
angular resolution is a major premium for imaging studies.  The spectral
energy distributions of brown dwarfs peak in the near-IR and hence these
wavelengths are advantageous for detection and characterization of
substellar binaries, especially for cooler objects.  Coincidentally,
ground-based adaptive optics (AO) systems are most easily developed at
these wavelengths.  Since the largest ground-based telescopes have
$\approx$4$\times$ larger apertures than \HST, and hence the potential
for $\approx$4$\times$ higher angular resolution, near-IR AO
observations are an appealing capability for examining substellar
multiplicity.  However, natural guide star AO observations are hampered
by the need for a bright star for wavefront sensing.  In particular, the
ultracool L and T~dwarfs are too faint for natural guide star AO, and
thus AO observations of these objects have largely been restricted to
the rare objects found as companions to bright stars
\citep{2001astro.ph.12407L, 2002ApJ...567L.133P, 2003ApJ...584..453F,
  2004A&A...413.1029M, 2004ApJ...617.1330M, 2005AJ....129.2849B}

Laser guide star (LGS) AO provides a powerful new tool for high angular
resolution imaging and spatially resolved spectroscopy of substellar
binaries.  Through resonant scattering off the sodium layer at
$\sim$90~km altitude in the Earth's atmosphere, sodium LGS systems
create an artificial star bright enough to serve as a wavefront reference
for AO correction \citep{1985A&A...152L..29F, 1987Natur.328..229T,
  1994OSAJ...11..263H}.  Thus, most of the sky can be made accessible to
near diffraction-limited IR imaging from the largest existing
telescopes.  Also, ground-based telescopes can serve as enduring
long-term platforms for high resolution imaging, which are needed for
dynamical mass determinations of substellar binaries.

Here we present the first brown dwarf science obtained with LGS AO,
observations of the ultracool dwarf Kelu-1 \citep{1997ApJ...491L.107R}.
Discovered in a photographic proper motion survey, this was one of the
first free-floating L~dwarfs identified.  Given its low effective
temperature, the presence of Li absorption in its spectrum clearly
demonstrated its substellar nature, based on the lithium test
\citep{1992ApJ...389L..83R,1998bdep.conf..394B}.  Kelu-1 has served as a
key object in spectral classification schemes for L~dwarfs
\citep{1999ApJ...519..802K, 1999AJ....118.2466M, geb01}.  It has been
assigned a spectral type of L2 by Kirkpatrick \etal\ based on optical
spectra and L3$\pm$1 by Geballe \etal\ based on optical and IR spectra.

Once its parallax was measured \citep{2000ASPC..212...74D}, Kelu-1
appeared to be overluminous compared to other early-L dwarfs
\citep{1999AJ....118.2466M, leg01}, attributed either to an unusually
young ($<$0.1~Gyr) age or an unresolved close companion.  As the number
of parallax measurements, and hence absolute magnitudes, for ultracool
dwarfs have grown, Kelu-1's apparent overluminosity has persisted
\citep{2002AJ....124.1170D, 2004AJ....127.2948V}.  Based on all the
observations to date, \citet{gol04} argued that unresolved binarity
remains the most compelling explanation, given the very young age
($\sim$10~Myr) otherwise needed to explain Kelu-1's variance with other
early-L~dwarfs in the solar neighborhood.  (Note that such a young age
would imply a model-derived mass of only $\sim$12~\Mjup\ for Kelu-1,
comparable to radial velocity planets found around other nearby stars.)

However, attempts to detect a companion to Kelu-1 have thus far been
unsuccessful.  \HST\ near-IR imaging failed to identify any companion as
close as 0.1\arcsec\ (1.9~AU; \citealp{1999Sci...283.1718M}), apparently
favoring the youthful interpretation of Kelu-1.
\citet{2002MNRAS.332..361C} detected photometric variability from Kelu-1
with a period of about 1.8~hrs, which they suggested might arise from
ellipsoidal variations due to a very close substellar companion, but
radial velocity monitoring data rule out this possibility
\citep{2003MNRAS.341..239C}.  Kelu-1's status as one of the archetypical
L~dwarfs provides a compelling reason to understand the origin of its
overluminosity.  Therefore, we targeted this object with the new LGS AO
system on the Keck~II Telescope as part of a high angular resolution
survey for substellar binaries.


\section{Observations}

We observed Kelu-1 on 2005~May~1~UT using the newly commissioned sodium
LGS AO system \citep{2004SPIE.5490..321B, 2004SPIE.5490....1W} of the
10-meter Keck II Telescope on Mauna Kea, Hawaii.  We used the facility
IR camera NIRC2 and the $J$ (1.25~\micron), $H$ (1.64~\micron), and
\Kp\ (2.12~\micron) filters from the Mauna Kea Observatories (MKO)
filter consortium \citep{mkofilters1, mkofilters2}.  Conditions were
photometric with excellent seeing conditions during the night, better
than 0.6\arcsec\ in the optical as reported by the neighboring Keck~I
Telescope.  The total setup time for the telescope to slew to Kelu-1 and
for the LGS AO system to be fully operational was 10~minutes.  

Kelu-1 was observed as it transited, at an airmass of 1.4.  The LGS
brightness, as measured by the flux incident on the AO wavefront sensor,
was equivalent to a $V\approx10.1$~mag star.  The LGS provided the
wavefront reference source for AO correction, with the exception of
tip-tilt motion.  This was sensed using an $R=14.3$~mag star from the
USNO-B catalog \citep{2003AJ....125..984M} located 27\arcsec\ away from
Kelu-1.  The LGS AO-corrected images have full widths at half maximum
(FWHM) of 0.08\arcsec, 0.07\arcsec, and 0.06\arcsec\ at $JH\Kp$,
respectively, with corresponding Strehl ratios of 0.03, 0.08, and 0.20.
The RMS variations in the FWHM and Strehl ratios between individual
images were about 6\% and 15\%, respectively.

We obtained a series of six images in each filter, dithering the
telescope by a few arcseconds between each pair of images.  The total
on-source integration time per filter was 3~minutes.  The sodium laser
beam was steered with each dither such that the LGS remained on Kelu-1
for all the images.
Kelu-1 was easily resolved into a binary system in all our data.  The
images were reduced in a standard fashion.  We constructed flat fields
from the differences of images of the telescope dome interior with and
without lamp illumination.  Then we created a master sky frame from the
median average of the bias-subtracted, flat-fielded images and
subtracted it from the individual images. Images were registered and
stacked to form a final mosaic (Figure~1).

To measure the flux ratios and relative positions of the two components,
we employed two approaches: (1) aperture photometry and (2) fitting of
an analytic model for the point spread function (PSF).  For aperture
photometry, we used small circular apertures ($\approx$1-3$\times$FWHM
diameter) centered on each component to determine the flux ratio,
measuring both the direct images and those created by removing the light
from the other component via rotating and subtracting the images.  For
the case of the analytic fit, the PSF was modeled as two gaussian
components, a narrow component for the PSF core and the broad component
for the PSF halo.  All the individual images were fit separately.  The
two methods agreed very well, as expected given the fact that the binary
is well-resolved and has a modest flux ratio.  We adopted the averages
of all the measurements of individual images as the final results and
the standard deviations as the errors.  For the separation and position
angle, no systematic offset was seen between the $JH\Kp$ dataset, and
thus we combined the measurements from all three filters.

Table~1 presents the resulting flux ratios and astrometry from the Keck
LGS AO images.  Table~2 reports the calculated IR colors and magnitudes
of the individual components.  To derive these from the observed flux
ratios, we used the integrated IR magnitudes reported in \citet{leg01}
and \citet{2004AJ....127.3553K}.
Any error due to photometric variability is likely to be negligible.
Kelu-1 shows variability at optical wavelengths with 1--2\% amplitude
\citep{2002MNRAS.332..361C, 2003MNRAS.341..239C}; near-IR variability at
this level would not impact our measurements.  Similarly, near-IR
monitoring of a small number of L~dwarfs finds little variability
\citep{2003MNRAS.339..477B, 2002A&A...389..963B}.  Absolute magnitudes
were determined using the parallactic distance of 18.7$\pm$0.7~pc
\citep{2002AJ....124.1170D}.\footnote{As described in \S~3, the
  projected separation of the binary has changed from 1998 to 2005.
  Therefore, the photocenter of the combined system has been moving
  relative to the center of mass (because each component's emitted flux
  is not linearly related to its mass).  In principle, such motion could
  have affected the parallax and proper motion determined by
  \citet{2002AJ....124.1170D}.  They obtained 30 observations over
  3.3~years, during which time the projected separation of the A~and B
  components may have changed by $\sim$0.15\arcsec.  This concern is
  slightly abated by comparison with the preliminary parallax reported
  by \citet{2000ASPC..212...74D}: their parallax based on 1.3~years
  worth of data is consistent with the 3.3-year dataset.  (However, the
  proper motion of the Dahn \etal\ measurements are somewhat discrepant
  with each other and with the results reported by Ruiz \etal\ 1997,
  which were based on two photographic observations taken 14~years
  apart.)  Once the orbit of the system is more fully determined, the
  relative motion of the photocenter can be accounted for in the
  astrometry calculations.}

Our NIRC2 $J$ and $H$-band photometry used the same MKO filters as
\citet{leg01} and \citet{2004AJ....127.3553K} and hence computing
magnitudes for the individual components from our measured flux ratios
is straight-forward.  However, our NIRC2 data were obtained with the MKO
\Kp-band (2.12~\micron) filter, in order to minimize the thermal
background from the sky + AO system, whereas the published photometry
used the MKO $K$-band (2.20~\micron) filter.  The measured near-IR
colors of L~dwarfs are sensitive to the specific choice of filters due
to the highly structured spectra of these very cool objects.  Following
\citet{2004PASP..116....9S}, we use synthetic photometry to determine
the (\Kp-$K$) color as a function of spectral type for L1 to T9 dwarfs:
\begin{equation}
(\Kp-K)_{MKO} = -0.007526 + 0.029263 \times (SpT)
                      - 0.0039505 \times (SpT)^2 
                      + 0.00010163 \times (SpT)^3
\end{equation}
where $SpT=0$ for L0 dwarfs, =1 for L1 dwarfs, =10 for T0 dwarfs, etc.
For the range of spectral types relevant to Kelu-1A and~B (see below),
the color term is 0.04--0.05~mags.  The RMS scatter about the polynomial
fit is 0.05~mags, which we add in quadrature when computing the $K$-band
photometric errors.

Aside from Kelu-1, the only other source detected in our images was
2MASS~J1305400$-$2541122, whose 2MASS $JH\Ks$ colors
\citep{1988PASP..100.1134B} and change in position relative to Kelu-1
between our 2005 images and the 1998 2MASS data are consistent with a
late-type background giant.
For objects at $\gtrsim$1\arcsec\ from Kelu-1, the final mosaics reached
a point source detection limit of 20.4, 20.5, and 19.9~mags at $JH\Kp$,
respectively, out to a separation of 3\arcsec.
Based on the models of \citet{2003A&A...402..701B} and an assumed age of
0.3--0.8~Gyr (see \S~4), these limits correspond to about
5--9~\Mjup\ companions around Kelu-1.


\section{Results}

The companion detected in our images has exceptionally red IR colors,
characteristic of L~dwarfs.  Also, the companion has redder colors than
the primary, indicating a lower temperature.  While our single epoch of
Keck data by itself cannot prove the two objects are physically
associated, it is highly unlikely the companion is an unrelated
background source.  For instance, in a 1~deg$^2$ region around Kelu-1,
the 2MASS catalog has only 2~other well-detected (S/N$>$10) sources of
comparable or fainter IR magnitudes and comparable or redder IR colors.
The odds of at least one such a source falling within our 100~arcsec$^2$
NIRC2 images are $1.5\times10^{-5}$.

Proof that the binary is a physically bound pair comes from the
\HST/NICMOS near-IR imaging obtained by \citet{1999Sci...283.1718M} in
August 1998.  They did not detect any companion to Kelu-1 as close as
0.1\arcsec\ separation (19~AU).  Retrieving their images from the HST
Archive, we find no other sources within 5\arcsec\ of Kelu-1.
Given Kelu-1's proper motion ($\mu$ = 0.285\arcsec/yr and
0.35\arcsec/yr, PA = 272\degs\ and 265\degs, as reported by
\citealp{2002AJ....124.1170D} and \citealp{1997ApJ...491L.107R},
respectively), the source we identify as Kelu-1B would be expected to
appear about 2\arcsec\ west of Kelu-1A in the NICMOS images if the two
objects were not physically associated.  Thus, the {\em non-detection}
of Kelu-1B in the earlier \HST\ imaging demonstrates that the system is
a physical pair (see \S~4.4 for further discussion).

To infer spectral types, we compared the measured $JHK$ colors and
absolute magnitudes of the two components to late-M and L~dwarfs with
known distances (Figure~2).  We assume that the components are
themselves single, and not unresolved binaries, an assumption that is
largely consistent with the observational constraints described below.
For the comparison field sample, we used the nearby ultracool dwarf
sample presented in \citet{leg01}, \citet{2004AJ....127.3553K}, and
\citet{gol04}, which have photometry on the MKO system and spectral
types using the \citet{geb01} scheme.  (We exclude the L2.5~dwarf SDSS
1435-0043, whose parallax is measured only with S/N=3, compared to the
S/N$\gtrsim$10 measurements typical for the rest of the sample.)  We add
objects from the \citet{2003ApJ...596..561M} spectral library with known
parallaxes, converting the magnitudes and spectral types to the Geballe
and MKO systems.

In general, $JHK$ colors alone are not well-suited for spectral typing
of L~dwarfs, since there is a large color scatter at a given type
\citep[e.g.][]{2000AJ....120..447K, leg01}.  The $JHK$ colors of Kelu-1A
and 1B are consistent with L1--L5 and L3--L9 spectral types.  However,
the integrated-light spectral type of L3 indicates that component~A
cannot be later-type than L3.  Furthermore, given this constraint on
Kelu-1A, the modest IR flux ratios (0.4--0.7~mags) mean that component~B
cannot be a very late-L~dwarf, since otherwise its magnitudes and
luminosity would be too faint.  If Kelu-1A is an L3~dwarf, Kelu-1B could
be no later than L6, based on the flux ratios between L3 dwarfs and
later-type L~dwarfs shown in Figure~2.\footnote{When using the observed
  $JHK$ flux ratios to estimate the spectral type of Kelu-1B relative to
  Kelu-1A, it may appear that we have implicitly assumed that the Kelu-1
  system has a similar age to the comparison field sample.  However,
  this is not the case.  We have only assumed that the flux ratios of
  different type L~dwarfs do not change, \eg, that the flux ratio of an
  early-L dwarf compared to a late-L dwarf is independent of age.  For a
  given conversion from spectral type to \Teff, this is equivalent to
  assuming that the ratio of the two radii is constant with age; this
  constancy is demonstrated by the fact that the isochrones in Figure~4
  are nearly parallel.}  Thus, the resolved $JHK$ colors, the published
integrated-light spectral type, and the IR flux ratios constrain the
spectral types to to be L1--L3 and L3--L6 for the two components.

Slightly more refined spectral type estimates come from comparing with
the absolute IR magnitudes of field objects.  IR magnitudes are
well-correlated with spectral type, except for the late-L's and
early-T's \citep{2002AJ....124.1170D, 2004AJ....127.2948V}.  We fit an
unweighted 3rd-order polynomial to the spectral type as a function of
absolute magnitude for field objects from M6 to L7.5.  We exclude known
binaries from the fit.\footnote{The close binaries plotted in Figure~2 are
  LHS~2397a (M8 primary; \citealp{2003ApJ...584..453F}),
  2MASSW~0746+2000 (L1 primary; \citealp{2001AJ....121..489R}),
  GJ~1001BC (L4 primary; \citealp{2004AJ....128.1733G}), DENIS-P
  J0204.4$-$1150 (L5.5 primary; \citealp{1999ApJ...526L..25K}), DENIS-P
  1228.2$-$1547 (L6 primary; \citealp{1999Sci...283.1718M}), 
  2MASSs~0850+1057 (L6 primary; \citealp{2001AJ....121..489R}), and 
  GL~337CD (L9.5 in combined-light; \citealp{2005AJ....129.2849B}).}  The
final resulting spectral type estimates for the two Kelu-1 components
are L1.5--L3 and L3--L4.5, with the independent $JHK$ datasets providing
consistent results.  The quoted range in spectral type arises from the
RMS scatter of the polynomial fit to the data.\footnote{Here we have
  assumed that the age of Kelu-1 is not very different than the
  comparison field sample, since we are comparing absolute magnitudes.
  However, as discussed in \S~4, Kelu-1 is somewhat younger than the
  field population but not enough to significantly affect the spectral
  type estimate (\eg, see the isochrones in Figure~4).}  With these
final spectral types, we compute the bolometric luminosities (\Lbol) of
each component, using the $K$-band bolometric corrections from
\citet{gol04}, and tabulate the results in Table~2.


\section{Discussion: Physical Properties in Light of the Binarity}

\subsection{Magnitudes, Colors, and \Teff}

The long-noted overluminosity of Kelu-1 relative to other early L~dwarfs
is clearly explained by its binarity.  The binarity also helps to
explain its position in IR color-magnitude diagrams (CMDs; Figure~2).
In integrated light, Kelu-1 appears to be unusually red in $J-K$
compared to its absolute magnitude, which led
\citet{2002AJ....124.1170D} to speculate on a possible connection
between its redness and its photometric variability
\citep{2002MNRAS.332..361C} and/or large $v \sin i$
\citep{2000ApJ...538..363B}.  However, Figure~2 shows that the resolved
photometry of the A and B components is consistent with the location of
other apparently single L~dwarfs.  (Interestingly, the location of
Kelu-1B in the CMDs is similar to known mid-L binaries, perhaps
suggesting that it is an unresolved system.)

Figure~2 also shows that nearly all the L~dwarf binaries are the reddest
objects in $J-K$ at a given magnitude.  This is expected: most known
substellar binaries have IR flux ratios close to unity, and since
later-type L~dwarfs are redder, unresolved L~dwarf binaries should
occupy a CMD position which is redder than other objects at the same
absolute magnitude.  Likewise, the most luminous sources at a given IR
color are the most likely candidates for binarity.\footnote{The clear
  exception in Figure~2 is GL~337CD, which has an L9.5 integrated-light
  spectral type on the Geballe \etal\ 2002) scheme (as computed by us
  using the spectrum from McLean \etal\ 2003).  This binary
  \citep{2005AJ....129.2849B} has an unusually blue $J-K$
  integrated-light color given its spectral type and IR magnitude.
  These properties could be explained if the binary is composed of a
  late-L dwarf and an early-T dwarf, given the comparable $J$-band
  magnitudes but greatly different IR colors of such objects.}
Therefore, some of the scatter in the IR CMD must arise from
unrecognized binarity.  Other effects can also introduce significant
scatter, such as a large age spread in the local population and the
effect of condensate clouds on the IR fluxes
\citep{2002ApJ...568..335M}.  A complete binary census of the nearby
L~dwarfs would be valuable to understand the relative significance of
these effects.


Binarity also explains the apparently discrepant effective temperature
(\Teff) of Kelu-1.  \citet{gol04} found that Kelu-1's \Teff\ of 2300~K
is $\approx$400~K hotter than the other L3~dwarfs in their sample
(accounting for the younger age of Kelu-1 relative to the rest of their
sample).  To estimate the \Teff\ of the individual components, we can
use the measured $K$-band flux ratio to estimate the bolometric
luminosity ratios, since Golimowski \etal\ find that the $K$-band
bolometric corrections are nearly constant for early to mid-L~dwarfs.
The individual \Teff's are then given simply by scaling from the
\Teff\ originally computed, \eg, $\Teff_{,A} = (L_A/L_{A+B})^{1/4}
\times (\Teff_{,A+B})$, where $L_A$ is the bolometric luminosity for
component~A and $L_{A+B}$ for the combined system.  Thus we estimate
$\Teff_{,A}\approx2020$~K and $\Teff_{,B}\approx1840$~K, in accord with
other L2--L3 and L4 dwarfs, respectively, studied by Golimowski \etal

\subsection{Lithium: Substellar Status and Age Determination}

Lithium is fully depleted over the lifetime of low-mass stars and high
mass ($\gtrsim$0.065~\Msun) brown dwarfs \citep{1992ApJ...389L..83R,
  1996ApJ...459L..91C, 1998ApJ...497..253U}.  The presence of Li
absorption in L dwarfs ($\Teff\approx1600-2300$~K; \citealp{gol04}) is a
clear indicator of substellar status, since objects with
\Teff$\lesssim$2700~K which show lithium must be below the
lithium-burning limit of 0.065~\Msun\ \citep{1998bdep.conf..394B}.
Kelu-1 shows Li~I~6708~\AA\ absorption in its spectrum, with an
equivalent width (EW) of about
1~\AA\ \citep{1997ApJ...491L.107R,1999ApJ...519..802K}.  However, the
binarity of Kelu-1 raises the question of whether only the lower-mass
component possesses Li, with the higher-mass component having destroyed
it.

A simple estimate shows that Kelu-1B could plausibly be the sole
Li-bearing component of the binary.  The flux ratios of L1--L2.5~dwarfs
(\ie, Kelu-1A) to L3--L4.5~dwarfs (\ie, Kelu-1B) are expected to be
$\approx$0.5--2.0~mags at optical wavelengths near the
Li~6708~\AA\ line, based on absolute magnitudes from
\citet{2002AJ....124.1170D} and \citet{2002astro.ph..4065H}.
If the Li absorption line resides solely with Kelu-1B, the broad-band
optical flux ratios allow us to estimate the dilution due to the Kelu-1A
and thus the true line strength.  (Here we ignore the detailed
differences between the optical continua of the early and mid-L~dwarfs.)
For optical flux ratios of 0.5--2.0~mags, Kelu-1B would have to have
EW(Li)$\approx$2.5--7~\AA\ to produce the observed combined-light
EW(Li).  Such an EW(Li) is consistent with other L3--L4 dwarfs
\citep{2000AJ....120..447K}.  Indeed, the combined-light EW(Li) of
Kelu-1 is relatively low compared to other early-L's, originally
suggestive of partial depletion \citep{1998ASPC..154.1819B,
  2000AJ....120..447K}.
This apparent deficiency is naturally explained by Kelu-1B being the
sole Li-bearing component.

However, given Kelu-1B's estimated spectral type of L3--L4.5
(\Teff$\approx$1750--1950~K; \citealp{gol04}), the fact that at least it
possesses lithium shows that {\em both} components of Kelu-1 are
substellar, regardless of whether Kelu-1A has Li.  This is shown in
Figure~3, which compares the derived \Lbol's for the two components
against models for lithium depletion from \citet{1997ApJ...491..856B}
and \citet{1998A&A...337..403B}.  The models indicate that the presence
of lithium gives an upper age of 0.8~Gyr for the system.  (We assume
that depletion of the initial abundance by a factor of~100 demarcates
the detection limit for lithium absorption.  Since depletion occurs
quickly, the derived upper age does not depend strongly on the adopted
depletion factor.)
This upper age of 0.8~Gyr results in upper mass estimates of
0.065--0.07~\Msun\ and 0.06-0.065~\Msun\ for components A and B,
respectively.  The stellar/substellar boundaries in the Burrows and
Baraffe \etal\ models are 0.075 and 0.072~\Msun, respectively, and
therefore we conclude that Kelu-1A is also a substellar object.

Spatially resolved high resolution optical spectroscopy of the two
components would refine the age estimate for Kelu-1 to better than is
generally possible for field objects.  Early-L dwarfs with ages of
$\gtrsim$0.6~Gyr are massive enough to deplete their lithium (Figure~3).
Thus, if Kelu-1A has no lithium but Kelu-1B does, the estimated age of
the system would be $\approx$0.6--0.8~Gyr.  Likewise, if both components
possess lithium, the binary's estimated age would be $\lesssim$0.6~Gyr.

Exploiting the differing Li depletion timescales for objects of
different masses offers an appealing method for age-dating L~dwarf
binaries.  This approach is analogous to the lithium depletion boundary
technique used to determine ages for young open clusters
\citep[e.g.][]{1996ApJ...458..600B}.  Age estimates are needed in
combination with dynamical masses to test theoretical evolutionary
models, but substellar binaries in the field have poorly constrained
ages.\footnote{For substellar binaries around main-sequence stars (see
  compilation in \citealp{2005AJ....129.2849B}), somewhat better
  constraints are possible based on indirect age indicators for the
  primary stars, such as stellar activity and/or kinematics
  \citep[e.g.][]{1999A&A...348..897L}.}
Li-bearing L~dwarf binaries are uncommon, with 6~such systems of the
21~known L~dwarf binaries (see compilation in
\citealp{2003ApJ...587..407C}, with the addition of GJ~1001BC from
\citealp{2004AJ....128.1733G} and Kelu-1AB from this work).  Among these
6 Li-bearing systems, 4~of them have separations larger than 0.2\arcsec,
making them amenable to resolved optical spectroscopy with very good
(\eg, tip-tilt corrected) angular resolution: DENIS-P~J1228.2$-$1547
\citep{1999Sci...283.1718M, 1997ApJ...490L..95T}, 2MASSW~J1239+5515
\citep{2003AJ....126.1526B, 2003AJ....125.3302G, 2000AJ....120..447K},
2MASSW~J1146+2230 \citep{2001AJ....121..489R, 1999ApJ...519..802K}, and
Kelu-1.
Measuring different lithium contents for their individual components
could provide a sample of binaries with age estimates better than
possible for typical substellar binaries.\footnote{The other two known
  Li-bearing L~dwarf binaries are the 0.06\arcsec\ system
  2MASSW~1112+3548 \citep{2003AJ....126.1526B, 2000AJ....120..447K} and
  the 0.16\arcsec\ system 2MASSs~J0850+1057
  \citep{2001AJ....121..489R,1999ApJ...519..802K} Note that 0850+10 may
  be composed of an L~dwarf primary and a T~dwarf secondary, in which
  case it may not be suitable for this age-dating approach since atomic
  lithium converts to molecular form for T~dwarfs
  \citep{1999ApJ...512..843B}. This method could also be applied to
  low-mass binaries composed of a late-M~dwarf and an L~dwarf.  However,
  since M~dwarfs deplete lithium much faster than L~dwarfs, the
  resulting age constraint would have a much larger range.}

\subsection{Estimated Age Range and Component Masses}

In order to estimate the masses of the components, we must first
estimate the age of the system.  The Li~detection discussed above leads
to an upper age limit of 0.8~Gyr.  The lower age is less
well-constrained but cannot be arbitrarily young: if the assumed age is
too small, then given the inferred \Teff, the observed magnitudes and
\Lbol's would be too bright compared to model predictions (see below).
From optical spectroscopy of gravity-sensitive features,
\citet{2003astro.ph..9634M} infer that Kelu-1 is older than the L2~dwarf
G196-3B, whose age is estimated to be 0.02--0.3~Gyr based on the
properties of its M~dwarf primary star G196-3
\citep{1998Sci...282.1309R}.
Hence, we initially adopt a lower bound of 0.1~Gyr for the Kelu-1 system.

As a first estimate of the masses, we use the measured \Lbol's with the
adopted 0.1--0.8~Gyr age and compare to theoretical models (Figure~3).
The resulting mass estimates are 0.025--0.07~\Msun\ and
0.02--0.065~\Msun\ for components A and~B, respectively.  The age of the
system dominates the uncertainties in the masses, since the iso-mass
lines in Figure~3 are mostly vertical.  Note that the model-inferred
mass ratio of the system is relatively independent of the exact age,
with component~A expected to be about 10--20\% more massive than
component~B.

A more refined estimate of the masses is possible by including the
effective temperatures as additional constraints.  We use our spectral
type estimates of L1.5--L3 and L3--L4.5 to derive \Teff\ for the two
components, based on the \citet{gol04} scale.  Figure~4 shows the
resulting mass estimates using the \Lbol, age, and \Teff\ information,
under the assumption that the conversion from spectral type to \Teff\ is
age-independent for $\gtrsim$0.1~Gyr.\footnote{The Golimowski
  \etal\ scale is derived from field objects, most of which are probably
  $\approx$10$\times$ older than Kelu-1AB.  Theoretical models predict
  that brown dwarf radii change by no more than 30\% from 0.1--10~Gyr
  \citep{bur01}.  The concomitant change in surface gravity is 0.2~dex,
  a small amount.  On the other hand, potentially important factors such
  as variations in dust content and rotation are not well-characterized
  and may lead to greater age-sensitivity in the relation between
  spectral type and \Teff\ than expected from surface gravity
  differences alone.}
The lower age range of 0.1~Gyr appears to set a lower mass estimate of
0.03~\Msun\ for Kelu-1B, depending on the choice of models.  However, if
the system were as young as 0.1~Gyr, Kelu-1A would be too faint for an
object of type L3 or earlier (\Teff$\gtrsim$1950~K) compared to the
models.  Thus, the \Teff\ and \Lbol\ constraints from Kelu-1A in
Figure~4 lead to a revised lower age limit of $\approx$0.3--0.4~Gyr,
depending on the choice of models, and thereby giving lower mass
estimates of 0.05~\Msun\ for~A and 0.045~\Msun\ for~B (Figure~3).  These
mass estimates are slightly more restrictive than those derived without
using the effective temperatures.

\subsection{Orbital Motion}

The joint information provided by the \HST\ non-detection and our Keck
detection can provide some crude insight into the binary's orbit, from
examining the chances that the \HST\ observation occurred at a time when
the binary's projected separation was much smaller than the current
value. \citet{2004ApJ...607.1003B} has computed the probability of
discovering a companion from a single observation in the situation where
the object cannot be detected too close to the primary, assuming a
random orbital phase and longitude of periastron.  This probability
depends on the companion orbital parameters and the minimum radius for
detection (the ``obscuration radius'').  Assuming that Kelu-1's orbit is
circular and that the 2005 position represents the extreme projected
separation (\ie, the semi-major axis), the ratio of the semi-major axis
to the \HST\ obscuration radius ($\approx$0.1\arcsec) is a factor of
$\approx$3.  \citet{2004ApJ...607.1003B} show that the probability of
not detecting such a companion in a single observation is about 5\%,
assuming a modest ($<$0.35) orbital eccentricity.  Thus under these
assumptions, the \HST\ non-detection was improbable but only at the
$\approx$2$\sigma$ level.  However, for a semi-major axis of 0.7\arcsec,
there is only a 1\% chance that \HST\ would not have seen the companion.
Therefore, the semi-major axis of Kelu-1B is unlikely to be greater than
$\sim$2$\times$ its current separation.

Basically, the true semi-major axis of the Kelu-1AB system is unlikely
to be much larger than the current projected separation, or else
\HST\ would have almost certainly resolved the binary.  Note that the
likelihood of the \HST\ non-detection rises if the orbit is eccentric
and the semi-major axis is closer to the obscuration radius, \ie,
smaller than the current separation \citep{2004ApJ...607.1003B}.  Also
if the orbit is circular, the \HST\ non-detection and Keck detection
indicate that the orbit must be close to edge-on, with a lower limit of
$\approx$70\degs\ for the inclination.  In this situation, the expected
amplitude of the radial velocity reflex motion would be $\approx$2~\kms,
perhaps detectable with current near-IR spectrographs provided
multi-year stability can be achieved.

We can make a rough estimate of the orbital period, assuming that the
true semi-major axis is not very different than the observed projected
separation.
For a 5.4~AU semi-major axis, the estimated total system mass of
$\approx$0.12~\Msun\ (0.095--0.135~\Msun) gives an expected orbital
period of 37~years (34--41~years).  More generally,
\citet{1999PASP..111..169T} show that $\approx$85\% of randomly oriented
orbits have a true semi-major axis of 0.5--2.0$\times$ the projected
separation, corresponding to periods of $\approx$15--105~yr for
Kelu-1AB.


\section{Summary}

Keck LGS AO imaging has revealed the binary nature of the nearby L~dwarf
Kelu-1.  Non-detection by \HST\ in 1998 demonstrates that the system is
a true physical pair, and that the orbital motion over the last 7~years
has been significant.  The non-detection also indicates that the true
semi-major axis of the system is unlikely to be much larger than the
currently observed 0.29\arcsec\ separation, or else the
\HST\ non-detection would have been highly improbable.

The binarity of Kelu-1 explains the long-noted anomalies of this object
relative to other early-L~dwarfs, namely its overluminosity, its very
red IR colors, and its unusually large inferred \Teff.  Its low
Li~absorption line strength relative to other early-L dwarfs may also be
explained by binarity --- Kelu-1B could be the sole component possessing
lithium, with the light from the non-Li-bearing Kelu-1A leading to a
relatively weak absorption line in the integrated-light spectrum.
Comparing the resolved $JHK$ colors and magnitudes, and the
integrated-light spectral type to other field objects leads to spectral
type estimates of L1.5--L3 and L3--L4.5 for the two components.
Published spectroscopy of lithium and gravity-sensitive features gives an
estimated age of 0.1--0.8~Gyr, and the resulting model-derived mass
estimates are 0.025--0.07~\Msun\ and 0.02--0.65~\Msun\ for Kelu-1A and
1B, respectively.  Assuming the relation between \Teff\ and spectral
type derived for field objects is appropriate for the somewhat younger
Kelu-1, the models provide a refined age estimate of 0.3--0.8~Gyr,
component masses of 0.05--0.07~\Msun\ and 0.045--0.065~\Msun, and a mass
ratio of about 0.9.

With an estimated orbital period of about 35~yrs ($\approx$15--105~yr
range), Kelu-1 joins the growing ranks of nearby L~dwarf binaries
potentially suitable for dynamical mass determinations.  Such
measurements would be a valuable test of substellar evolutionary models.
In addition, there are only six Li-bearing L~dwarf binaries known so
far, including Kelu-1AB; dynamical masses for these systems would
directly test the model-predicted $\approx$0.065~\Msun\ limit for
lithium burning.  Kelu-1 is one of four known Li-bearing L~binaries with
projected separations greater than 0.2\arcsec.  For these systems,
determining whether one or both components possess lithium could provide
high precision constraints on their ages, via the differences in the
lithium depletion timescale of the individual components.

Our Kelu-1 study represents the first LGS AO observations of a brown
dwarf.  But it provides only a small indication of what will be possible
as the Keck LGS system matures and other 8-10-meter class telescopes
deploy LGS AO systems.  Given the importance of high angular resolution
near-IR observations for substellar astronomy, our work is surely only
the first of many studies arising from this new capability.


\acknowledgments

We gratefully acknowledge the Keck LGS AO team for their impressive
efforts in bringing the LGS AO system to fruition.  It is a pleasure to
thank Antonin Bouchez, David LeMignant, Marcos van Dam, Randy Campbell,
Gary Punawai, Peter Wizinowich, and the Keck Observatory staff for their
assistance with the observations; Adam Burrows and Isabelle Baraffe for
providing model calculations; Neill Reid and Mike Cushing for insightful
discussions; and John Rayner, Dave Golimowski, and Gibor Basri for
comments on the manuscript.  Our research has benefitted from the
\HST\ Archive; the 2MASS data products; the SIMBAD database operated at
CDS, Strasbourg, France; and the M, L, and T dwarf compendium housed at
DwarfArchives.org and maintained by Chris Gelino, Davy Kirkpatrick, and
Adam Burgasser \citep{2003IAUS..211..189K, 2004AAS...205.1113G}.  This
work has been supported by NSF grant AST-0507833 and the Alfred P. Sloan
Foundation.  We thank Alan Stockton for a fortuitous swap of observing
nights.  Finally, the authors wish to recognize and acknowledge the very
significant cultural role and reverence that the summit of Mauna Kea has
always had within the indigenous Hawaiian community. We are most
fortunate to have the opportunity to conduct observations from this
mountain.

\clearpage


\begin{figure}
\vskip -8in
\vskip 7in
\hskip 2.1in
\centerline{\includegraphics[width=6in,angle=90]{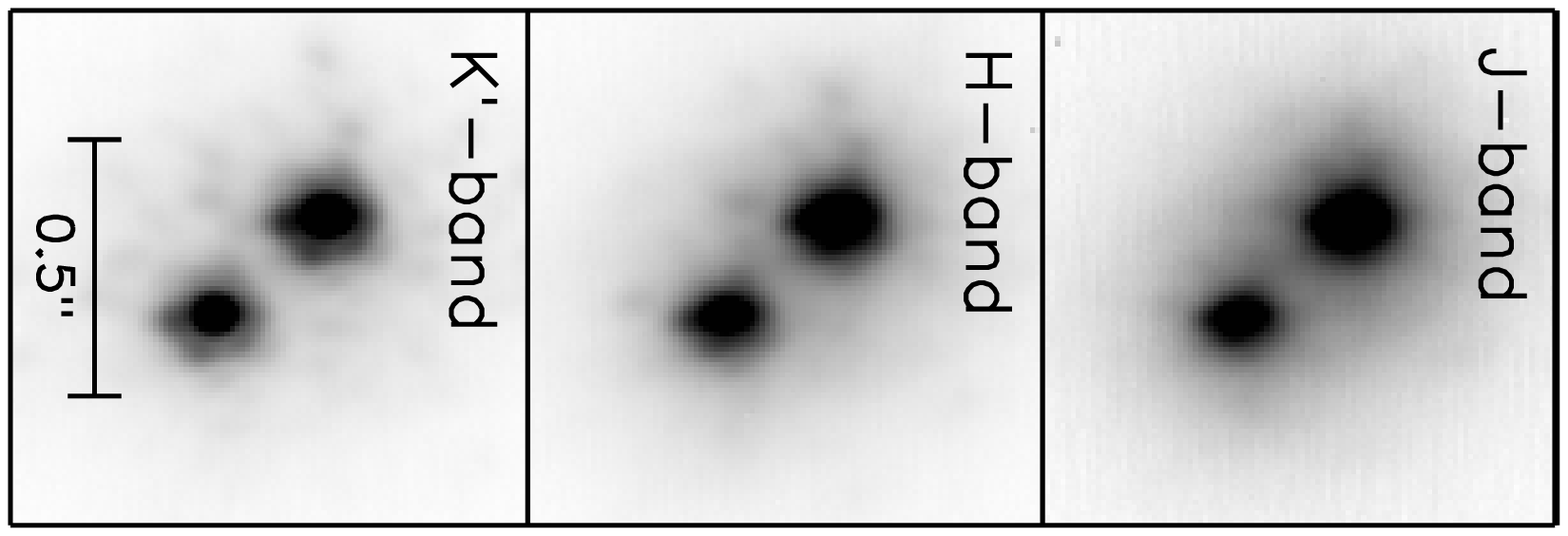}}
\vskip -6.02in
\hskip 4.2in
\centerline{\includegraphics[width=6in,angle=90]{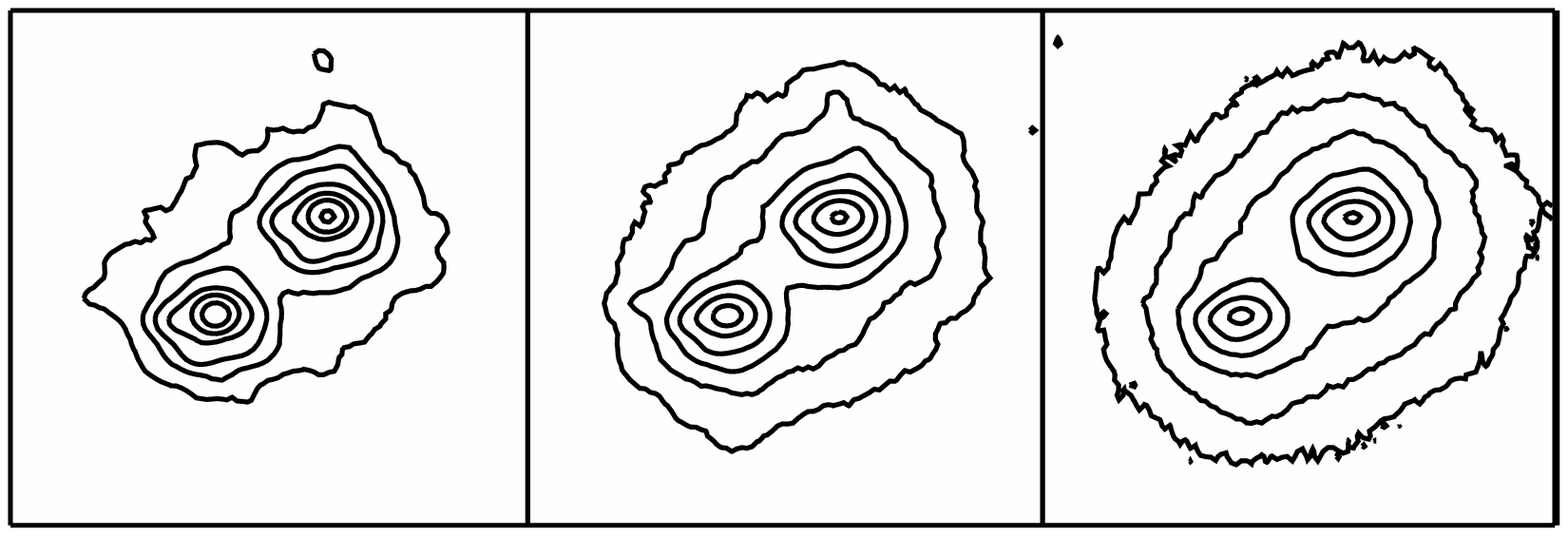}}
\vskip 1ex
\caption{\normalsize $JH\Kp$-band imaging of Kelu-1 from Keck LGS AO.
  North is up and east is left.  Each image is 1.0\arcsec\ (18.7~AU) on
  a side. The binary separation is 0.291\arcsec\ $\pm$
  0.002\arcsec\ with a position angle of 221.2\degs\ $\pm$
  0.6\degs\ east of north.  
  The contours are drawn from 90.0, 45.0, 22.5, 11.2, 5.6, 2.8, and
  1.4\% of the peak value in each bandpass.  The central region of the
  PSF is slightly elongated in the vertical direction due to the
  combined effects of atmospheric dispersion (more apparent at the
  shorter wavelengths) and slight asymmetries in the first Airy ring
  (more apparent at the longer wavelengths).}
\end{figure}

\begin{figure}
\vskip -0.25in
\centerline{\includegraphics[width=5in,angle=0]{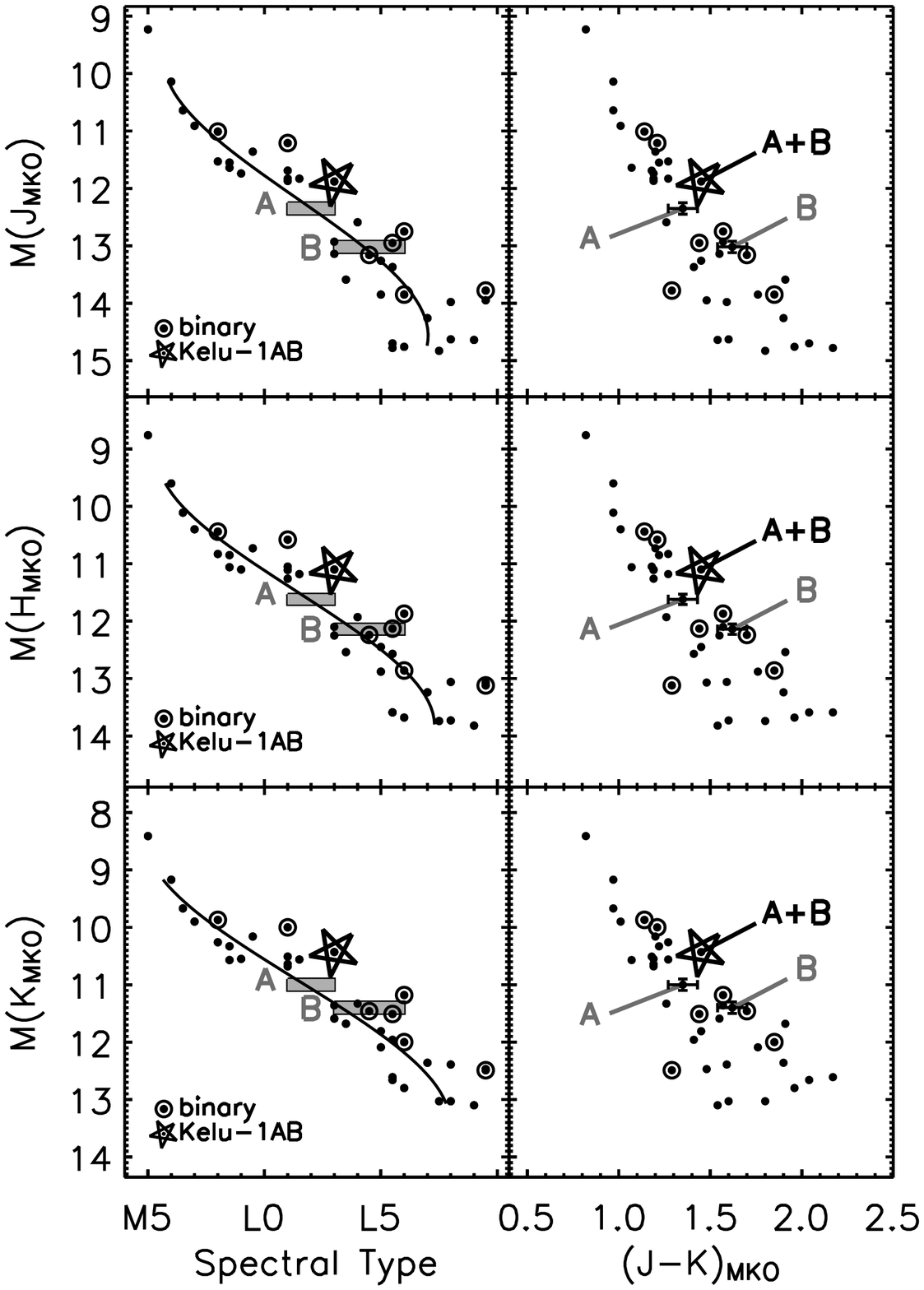}}
\vskip 1ex
\caption{\normalsize Near-IR properties of Kelu-1A and~B compared with
  nearby late-M dwarfs and L~dwarfs (see \S~3).  Integrated-light data
  for known binaries are shown as ringed dots, with Kelu-1AB shown as a
  star.  {\bf Left:} $JHK$ absolute magnitudes versus spectral types on
  the \citet{geb01} system.  The observational constraints from the
  absolute magnitudes and the spectral type range inferred from $JHK$
  colors, flux ratios, and integrated-light spectra are shown as shaded
  boxes.  The solid lines are fits to the spectral type as a function of
  absolute magnitude for the nearby dwarfs, excluding known binaries.
  The final inferred spectral types are L1.5--L3 for Kelu-1A and
  L3--L4.5 for Kelu-1B.  {\bf Right:} Infrared color-magnitude diagrams.
  In integrated light, Kelu-1AB appears to be unusually red for its
  absolute magnitude.  However, the resolved properties of the two
  components (shown as dots with error bars) are consistent with other
  apparently single L~dwarfs.
 \label{plot-sptype}}
\end{figure}

\begin{figure}
\centerline{\includegraphics[width=4.5in,angle=0]{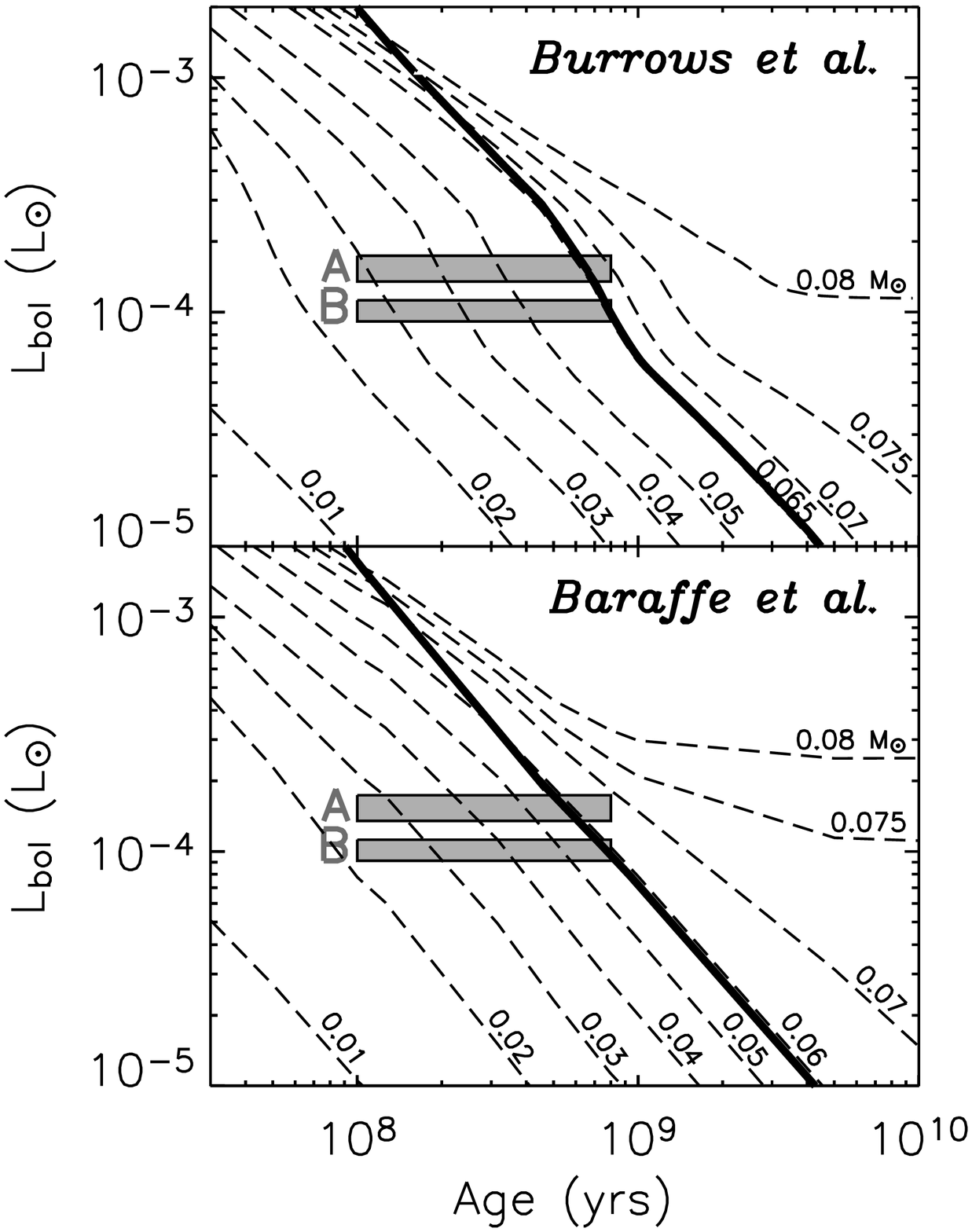}}
\vskip 6ex
\caption{\normalsize Mass estimates for Kelu-1A and 1B derived from
  components' inferred \Lbol\ and age, using Burrows \etal\ (1997) and
  \citet{1998A&A...337..403B,2003A&A...402..701B} models.  The
  stellar/substellar boundaries for the two sets of models are at
  $\approx$0.075~\Msun\ and $\approx$0.072~\Msun, respectively.  Dashed
  lines show models of constant mass, labeled in units of \Msun.  The
  heavy solid line represents the 1\% lithium depletion boundary;
  objects to the right of the line have depleted their lithium.  The
  shaded rectangles show the observational constraints for Kelu-1A
  and~1B.  The estimated age range is 0.1--0.8~Gyr, based on optical
  spectroscopy (see~\S~4).  Using the Burrows models, the resulting mass
  estimates are 0.03--0.07~\Msun\ and 0.025--0.065~\Msun\ for components
  A and~B, respectively.  The Baraffe models predict masses of
  0.025--0.07 and 0.02-0.06~\Msun.  Since the iso-mass lines are mostly
  vertical in this plot, the uncertainties in the masses are dominated
  by the uncertain age of the system.}
\end{figure}

\begin{figure}
\includegraphics[width=3in,angle=0]{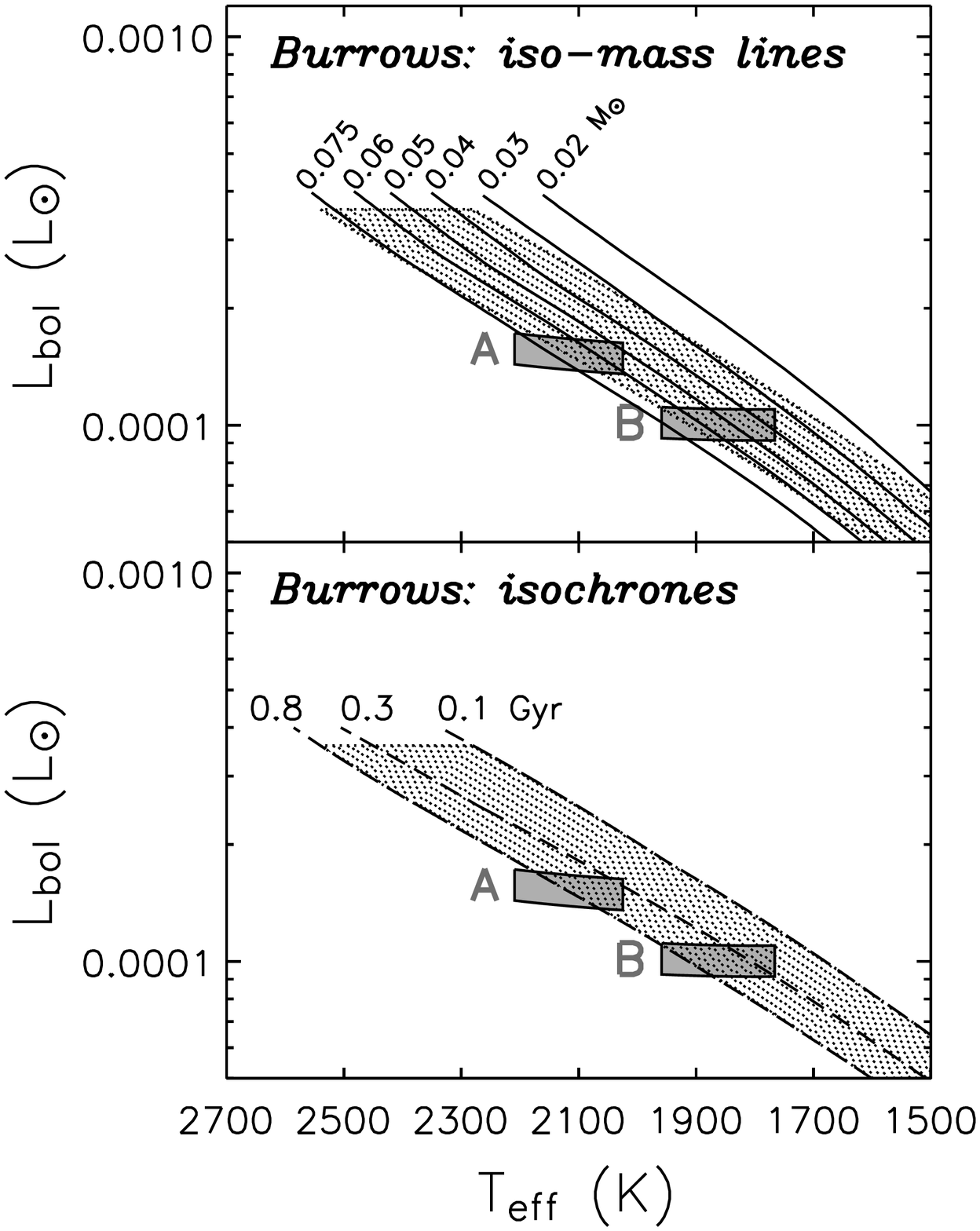}
\hskip 0.5in
\includegraphics[width=3in,angle=0]{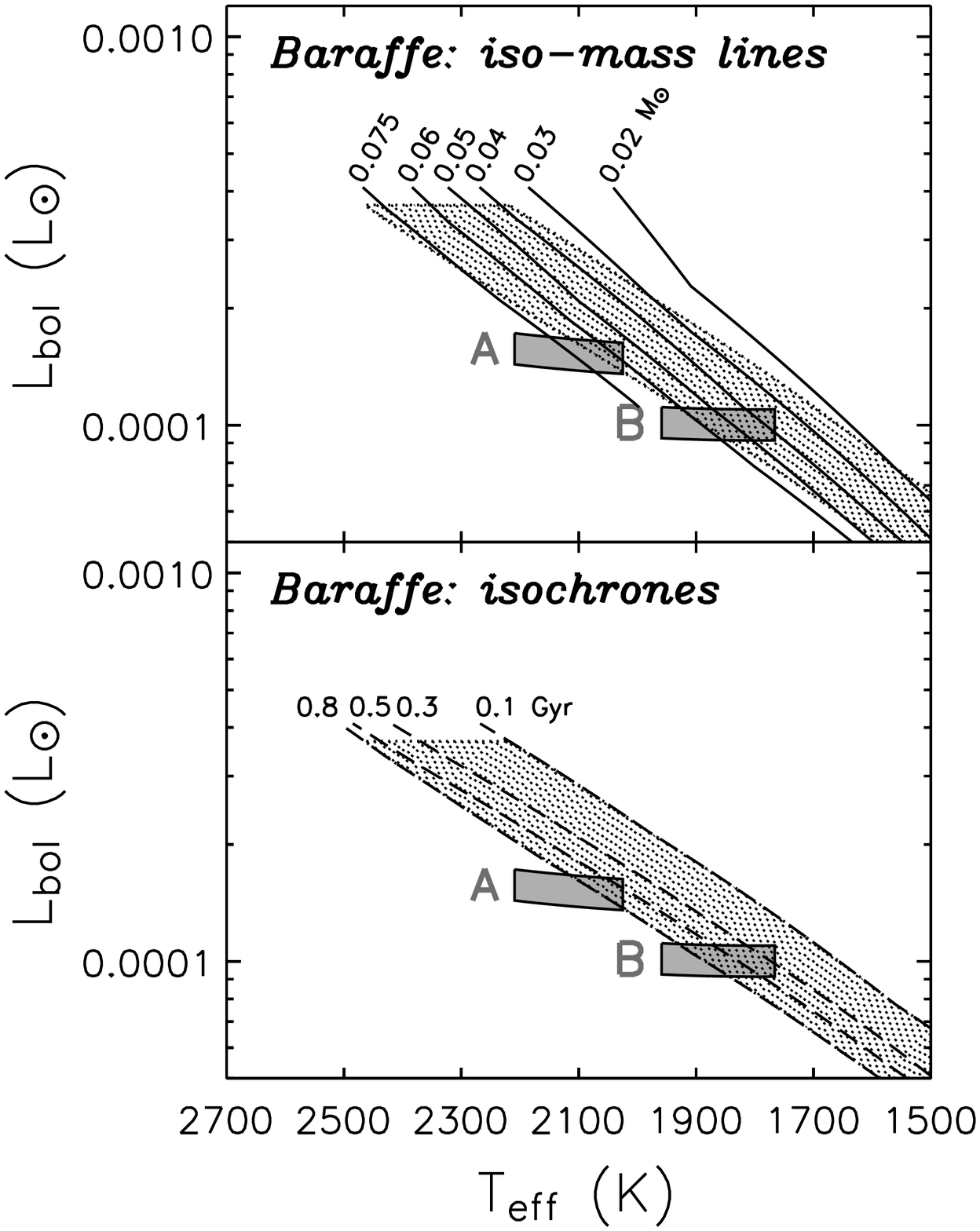}
\vskip 3ex
\caption{\normalsize Refined mass estimates for Kelu-1A and 1B, based on
  the estimated \Teff's combined with the constraints in Figure~3 (see
  \S~4.3).  The dark shaded horizontal regions show the \Teff\ and
  \Lbol\ range of the two components.  (Note that the \citealp{gol04}
  bolometric corrections depend on the spectral type, and hence the
  \{\Teff, \Lbol\} constraints are curved regions rather than
  rectangles.)  The dotted region shows the estimated age range of
  0.1--0.8~Gyr (truncated at the high luminosity end for plotting
  purposes).  Thus, the intersection of the dotted and the shaded
  regions represents all the available observational constraints.  {\bf
    Top plots:} Model tracks of constant mass are overplotted.  The
  inferred upper mass estimates are 0.07 and 0.065~\Msun\ for A and B,
  respectively, from the Burrows models.  The Baraffe models give values
  of 0.07 and 0.06~\Msun.  {\bf Bottom plots:} Same data and models, now
  with only model isochrones overplotted for clarity.  (The isochrones
  in fact are nearly parallel to the iso-mass lines.)  The observed
  \{\Teff, \Lbol\} of Kelu-1A suggest that a lower age limit of
  $\approx$0.3--0.4~Gyr, depending on the choice of models.  This leads
  to lower mass estimates of 0.05 and 0.045~\Msun\ for the two
  components.}
\end{figure}





\clearpage

\begin{deluxetable}{lc}
\tablecaption{Keck LGS AO Results\tablenotemark{a} \label{sample}}
\tablewidth{0pt}
\tablehead{
  \colhead{Property} &
  \colhead{Measurement} 
}

\startdata

$a$ (mas)            &  291 $\pm$ 2 \\
$\phi$ (deg)         &  221.2 $\pm$ 0.6 \\
$\Delta{J}$ (mags)   &  0.67 $\pm$ 0.04 \\
$\Delta{H}$ (mags)   &  0.52 $\pm$ 0.03 \\
$\Delta{\Kp}$ (mags) &  0.40 $\pm$ 0.02 \\

\enddata

\tablenotetext{a}{All photometry on the MKO system.}

\end{deluxetable}

\clearpage

\begin{deluxetable}{lcc}
\tablecaption{Resolved Properties of Kelu-1\tablenotemark{a}}
\tablewidth{0pt}
\tablehead{
  \colhead{Property} &
  \colhead{Kelu-1A} &
  \colhead{Kelu-1B}
}

\startdata
$J$ (mags)         &  13.70 $\pm$ 0.06  &  14.37 $\pm$ 0.06  \\
$H$ (mags)         &  12.97 $\pm$ 0.05  &  13.49 $\pm$ 0.05  \\
$K$ (mags)         &  12.35 $\pm$ 0.06  &  12.75 $\pm$ 0.06  \\

$J-H$ (mags)       &   0.73 $\pm$ 0.08  &   0.88 $\pm$ 0.08  \\
$H-K$ (mags)       &   0.62 $\pm$ 0.08  &   0.74 $\pm$ 0.08  \\
$J-K$ (mags)       &   1.35 $\pm$ 0.08  &   1.62 $\pm$ 0.08  \\

$M(J)$ (mags)      &  12.35 $\pm$ 0.10  &  13.02 $\pm$ 0.10  \\
$M(H)$ (mags)      &  11.62 $\pm$ 0.09  &  12.14 $\pm$ 0.09  \\
$M(K)$ (mags)      &  11.00 $\pm$ 0.10  &  11.40 $\pm$ 0.10  \\

Estimated spectral type &     L1.5 -- L3     &  L3 -- L4.5        \\
log(\Lbol/\Lsun)   & $-$3.76 to $-$3.87 & $-$3.95 to $-$4.04 \\
\enddata

\tablenotetext{a}{All photometry on the MKO photometric system.}

\end{deluxetable}

\end{document}